\documentstyle[a4,12pt]{article}
\begin{document}
\begin{titlepage}
\renewcommand{\thefootnote}{\fnsymbol{footnote}}
\rightline{DAMTP-R92/28}
\vspace{0.8in}
\LARGE
\center{Model Dependence of Baryon Decay Enhancement by Cosmic Strings}
\Large
\vspace{0.8in}
\center{C.J. Fewster\footnote{E-mail address: cjf10@uk.ac.cam.phx} and B.S.
Kay\footnote{Address from 1 October 1992: Department of Mathematics, University
of York, Heslington, York YO1 5DD, U.K.}
\setcounter{footnote}{0}
\renewcommand{\thefootnote}{\arabic{footnote}}
\vspace{0.5in}
\large
\center{\em Department of Applied Mathematics and Theoretical Physics,
\\ University of Cambridge, \\  Silver Street, Cambridge CB3 9EW, U.K.}}
\vspace{0.3in}
\center{July, 1992}
\vspace{1.0in}
\begin{abstract}
Cosmic strings arising from GUTs can catalyse baryon decay processes with
strong interaction cross sections. We examine the mechanism by which the cross
section is enhanced and find that it depends strongly on the details of the
distribution of gauge fields within the string core. We propose a calculational
scheme for estimating wavefunction amplification factors and also a physical
understanding of the nature of the enhancement process.
\end{abstract}
\end{titlepage}
\input mssymb
\newcommand{\Avec}{{\bf A}}
\newcommand{\Bvec}{{\bf B}}
\newcommand{\zvec}{{\bf \hat{z}}}
\newcommand{\nvec}{{\bf n}}
\newcommand{\thvec}{\bf \hat{\hbox{\boldmath $\theta$}}}

\section{Introduction}

Grand unified theories (GUTs) predict a rich variety of topologically stable
``defects'' -- domain walls, monopoles, cosmic strings -- whose localized
concentrations of unbroken gauge fields and Higgs condensate would be expected
to catalyse baryon decay.  In  particular, there would be a non-zero amplitude
for quarks which penetrate the core of such a defect to decay into leptons, as
quarks and leptons appear in the same multiplet in GUTs. The typical size of
such defects (monopole radius, or cosmic string radius) will be of the order of
the GUT length scale ($10^{-30}$cm). Naively, one might expect the cross
section for monopole catalysed baryon decay processes to be of the order of the
corresponding area ($10^{-60}\hbox{cm}^2$) and that for decay processes
mediated by a string, the cross section per unit length of string  would be of
the order of the GUT length.  In the case of monopoles however, it has long
been understood, since the work of Callan and Rubakov \cite{Cal,Rub}, that
cross sections for such processes will be greatly enhanced and actually be of
the order of the QCD area ($10^{-30}\hbox{cm}^2$). The essential point is that
there is a mechanism (due to the long range external magnetic field of the
monopole) for the amplification of the quark wavefunction at the monopole core
and thereby an increased probability of penetration. Since quark masses and
energies will be of the order of the QCD scale which is typically 15 orders of
magnitude smaller than the inverse radius of the monopole, one is thus
interested in the {\it low} energy scattering of quarks on monopole targets.
(We shall, however, treat energies sufficiently high that quarks may be treated
as free.)

In the case of cosmic strings, mechanisms for a similar enhancement of baryon
catalysis have often been discussed \cite{AW,AWMR,Anne}. Alford and Wilczek
\cite{AW} showed that a GUT string can carry fractional flux in units of
$2\pi/q$, where $q$ is the charge of a quark or lepton in the theory. Such a
particle thus has a significant low energy elastic cross section for scattering
from the string due to the presence of the topologically non-trivial, but pure
gauge external field configuration, as shown long ago by Aharonov and Bohm
\cite{AB}. It was also noted in \cite{AW} that the fermion wavefunction is
amplified at the string. Physically, this means that the presence of the
external gauge fields has increased the probability of fermions penetrating the
region of unbroken symmetry, thereby enhancing the catalysis rate.

In this paper, we follow the computational scheme for catalysis cross sections
originally proposed in \cite{Anne} which breaks the calculation into two steps.
In the first, the decay cross section is computed using free fermions (i.e.
ignoring the external gauge fields) to give a GUT scale cross section, known as
the {\em geometric cross section}. This step clearly depends strongly on the
decay mechanism -- whether baryon decay is mediated by interactions with
internal X and Y gauge fields or by interaction with a scalar condensate in the
string core. This is partly determined by the model of the string used. In the
second step, one computes the degree to which the geometric cross section is
enhanced by considering the scattering of fermions off the string. The
prescription used is derived from first order perturbation theory and gives the
decay cross section as
\begin{equation}
\frac{d\sigma}{d\Omega} = A^4\left.\frac{d\sigma}{d\Omega}\right|_{\rm geom}
\end{equation}
where $(d\sigma/d\Omega)_{\rm geom}$ is the geometric cross section and $A$ is
the {\em wavefunction amplification factor}
\begin{equation}
A= \frac{|\psi(a)|}{|\psi_{\rm free}(a)|} \label{eq:amplif}
\end{equation}
i.e. the ratio of the magnitude of the spinor in the magnetic field to the free
spinor evaluated at $a$ the core radius of the string (at infinity, both
spinors are normalised by scattering boundary conditions).

It was shown in \cite{Anne} that the total cross section can be enhanced up to
the QCD scale. The second step also depends on the internal model of the string
(gauge fields or scalar condensate) and also on the net flux carried by the
string; however, it was concluded in \cite{Anne} that the {\it distribution} of
gauge fields within the core does not affect the amplification factor. This
claim was made on the basis of a consideration of two simple models of flux
distribution: the case of a flux ring at the core radius and the case of
uniformly distributed flux within the core.

Here, we re-examine this claim by considering more general models of the flux
distribution. We shall concentrate exclusively on the case where the dominant
low energy scattering of fermions off the string is due to interactions with
gauge fields and where the decay process itself is mediated by gauge fields,
although our general methodology could easily be extended to cover interactions
with scalar fields. Under certain assumptions, we demonstrate that the results
of \cite{Anne} are indeed independent of the details of the flux distribution.
However, when the flux is allowed e.g. to change sign in the string interior,
we
find that the amplification factors can be strongly dependent on the details of
the flux distribution. It might be objected that such a field configuration is
unphysical, and certainly a single gauge field whose flux changed sign in the
core would probably be unstable. However, the situation we envisage is where
two or more gauge fields may be represented by a single {\it effective} $U(1)$
gauge field. For example, these fields might be the X and Y gauge fields and
the electromagnetic field. If the separate gauge fields have different ranges
(as they do in this example) and if they are of opposite, but constant sign
(and there is no {\it a priori} reason to prevent this) then the effective
gauge field could certainly change sign in the core without prejudicing the
stability of the string. Thus our results may be important for computations
with realistic string models.

In order to treat these problems, we develop and extend a calculational scheme
(the scattering length formalism) which was originally developed \cite{KF,FK}
(see also \cite{KS}) to study the general problem of the large scale behaviour
of small objects. A remarkable property of Dirac operators coupled to external
$U(1)$ gauge fields allows us to calculate the relevant parameters (the
scattering lengths introduced below) analytically for {\it arbitrary} flux
distributions and by developing the analogue to the low energy expansion of
potential scattering theory \cite{Newt} we are thus able to provide a simple
means of estimating the amplification factors for baryon decay. Our formalism
leads to a physical interpretation of the enhancement process. We also comment
on the validity of the thin wire approximation to the flux distribution used in
\cite{AW,AWMR} and relate our results to our other work on the large scale
effects of small objects \cite{KF,FK}.

Our principal assumptions are as follows:
\begin{enumerate}
\item Quarks are treated (as in \cite{AWMR,Anne}) as free Dirac
particles with energies above the confinement scale. (On the GUT scale, this
still corresponds to low energies.)
\item Decay processes are mediated by interactions with gauge fields in the
core.
\item The cosmic string is assumed to be an infinitely long straight
cylindrical string along the $z$-axis of radius $a$.
\item All fields are cylindrically symmetric about the $z$-axis.
\item The quark wavefunction is $z$-translationally invariant. (Due to the low
energies of incoming quarks, we expect that including the $z$-dependence of the
quark wavefunction will not significantly alter the physics.)
\end{enumerate}

  Our conventions are as follows: the metric has signature $+---$, the incoming
quark has charge $-e$ and the electromagnetic vector potential is defined by
$\hbox{$A_{\mu}= (\phi,-\Avec)$}$ with $\nabla\wedge\Avec=\Bvec$, where $\Bvec$
is the magnetic field\footnote{We refer to $A_{\mu}$ in terms appropriate to an
electromagnetic field; however it should of course be thought of as the
effective gauge field.}. $(r,\theta,z)$ are cylindrical polar coordinates about
the $z$-axis.

\section{The Dirac Equation with a Flux Tube}

 We consider the minimally coupled Dirac equation
$(i\gamma^{\mu}(\partial_{\mu}-ieA_{\mu})-\tilde{m})\psi_4=0$
in the $\gamma$-matrix representation
\[
\gamma^0=\left(\begin{array}{cc}\sigma_3 & 0\\ 0 & -\sigma_3\end{array}\right),
\gamma^1=\left(\begin{array}{cc}i\sigma_2 & 0\\ 0 &
-i\sigma_2\end{array}\right),
\gamma^2=\left(\begin{array}{cc}-i\sigma_1 & 0 \\ 0 &
i\sigma_1\end{array}\right),
\gamma^3=\left(\begin{array}{cc} 0 & 1 \\ -1 & 0 \end{array} \right).
\]

For a flux tube of magnetic flux $\Bvec = B(r)\zvec$ with $B(r)$
vanishing for $r>a$, a simple Stokes' theorem argument gives
\[ \Avec = \frac{\alpha(r)}{er}\thvec \]
where $\alpha(r)$ is defined by
\[ \alpha(r) = e\int_0^r B(r^{\prime})r^{\prime} dr^{\prime}. \]
We define $\Phi=\alpha(a)\equiv (\hbox{total magnetic flux})/(2\pi/e)$. In
addition, we define $\nu=\Phi-[\Phi]$, where $[\Phi]$ is the greatest integer
{\it strictly} less than $\Phi$. We will only be interested in the case of
non-integer flux.

Diagonalising $J_z$, the angular momentum operator about the $z$-axis, given by
\begin{equation}
 J_z = -i\partial_{\theta}+\frac{1}{2}\left(\begin{array}{cc} \sigma_3 & 0 \\
                                                          0 & \sigma_3
                                            \end{array} \right)
\end{equation}
with eigenvalues $n+\frac{1}{2}$ ($n\in{\Bbb Z}$) using the {\it ansatz}
\begin{equation}
\psi_4 = e^{-i\omega t}\exp \left\{ i\left[ n+\frac{1}{2}-
\frac{1}{2}\left(\begin{array}{cc} \sigma_3 & 0 \\
                                   0 & \sigma_3
                 \end{array} \right) \right] \theta \right\}
           \left(\begin{array}{c} F_n^{\uparrow}(r) \\ G_n^{\uparrow}(r) \\
                                     F_n^{\downarrow}(r) \\ G_n^{\downarrow}(r)
                 \end{array} \right)
\end{equation}
we find that the Dirac equation separates into two 2-spinor equations for
$F_n^{\uparrow},G_n^{\uparrow}$ and $F_n^{\downarrow},G_n^{\downarrow}$ with
radial equations
\begin{equation}
\left.\begin{array}{ccc}
-i(\omega+m)G_n+\left(\frac{d}{dr}-\frac{n+\alpha(r)}{r}\right)F_n & = & 0 \\
-i(\omega-m)F_n+\left(\frac{d}{dr}+\frac{n+1+\alpha(r)}{r}\right)G_n & = & 0
\label{eq:coup}
\end{array} \right\}
\end{equation}
where $m=\tilde{m}$ for $F_n^{\uparrow},G_n^{\uparrow}$ and $m=-\tilde{m}$ for
$F_n^{\downarrow},G_n^{\downarrow}$.

 In the rest frame of the particle, $\uparrow$ and $\downarrow$ correspond to
spin aligned and anti-aligned with $\zvec$. We now drop the arrows and proceed
to treat only one 2-spinor in each angular momentum sector:
\begin{equation}
\psi_n = \left( \begin{array}{c} F_n(r) \\ G_n(r) \end{array} \right).
\end{equation}
It will turn out that the amplification factor has the same order of magnitude
whichever 2-spinor is chosen.

The equations~(\ref{eq:coup}) decouple to give
\begin{eqnarray}
\left\{ -\frac{1}{r}\frac{d}{dr}r\frac{d}{dr}
+V_n^+
+m^2-\omega^2\right\}F_n = 0                    \label{eq:uptrue} \\
\left\{ -\frac{1}{r}\frac{d}{dr}r\frac{d}{dr}
+V_n^-+m^2-\omega^2\right\}G_n = 0              \label{eq:dntrue}
\end{eqnarray}
where the effective potentials are given by
\begin{eqnarray}
V_n^+=\left(\frac{n+\alpha(r)}{r}\right)^2+\frac{\alpha^{\prime}(r)}{r}
\label{eq:Vplu} \\
V_n^-=\left(\frac{n+1+\alpha(r)}{r}\right)^2-
\frac{\alpha^{\prime}(r)}{r}.        \label{eq:Vmin}
\end{eqnarray}

  In the subsequent discussion we shall see that an understanding of the
mechanism for enhancement depends crucially on a consideration of the details
of the effective potentials $V_n^{\pm}$ (i.e. on the details of the flux
distribution) inside the cosmic string, and that one loses essential insight
if one passes to the thin wire approximation, or if one takes non-smooth models
for the flux distribution.

Equations~(\ref{eq:uptrue}) and~(\ref{eq:dntrue}) are not independent, but are
coupled via the first order equations~(\ref{eq:coup}). It is easy to see that
the coupling ensures that if one component is regular at the origin, then so is
the other. Solving the system at momentum $k=\sqrt{\omega^2-\tilde{m}^2}$ with
regular boundary conditions at $r=0$, we find that outside the string core,
where $\alpha(r)\equiv\Phi$, $F_{n,k}(r)$ and $G_{n,k}(r)$ may be written as
\begin{eqnarray}
F_{n,k}(r)\propto
\cos\theta_n J_{|n+\Phi|}(kr) - \sin\theta_n J_{-|n+\Phi|}(kr)
\label{eq:ext1} \\
G_{n,k}(r)\propto
\cos\varphi_n J_{|n+\Phi+1|}(kr) -\sin\varphi_n J_{-|n+\Phi+1|}(kr)
\label{eq:ext2}
\end{eqnarray}
where $\theta_n$ and $\varphi_n$ must be determined by matching the external
solution to the solution inside the string core. The coupling between
equations~(\ref{eq:coup}) implies that $\tan\theta_n=-\tan\varphi_n$ in all
sectors except $n=-1-[\Phi]$, where we require
$\tan\theta_n=-\cot\varphi_n$.

If we were to employ the ``thin wire approximation'', in which the flux is
concentrated in a flux tube of infinitesimal radius, the external
solutions~(\ref{eq:ext1}),(\ref{eq:ext2}) would hold down to $r=0$. We could
then apply the criterion of local square integrability (with measure $rdr$) at
$r=0$ to fix $\theta_n=\varphi_n=0$ (for all $k$) in all sectors other than
$n=-1-[\Phi]$, which we refer to as the {\it critical sector}. In the critical
sector any choice of $\theta_n$ and $\varphi_n$ consistent with the coupling
$\tan\theta_n=-\cot\varphi_n$ leads to a locally square integrable
wavefunction. We therefore see that in the thin wire approximation, all
wavefunctions are regular and vanishing at the origin except in the critical
sector where there is a 1-parameter family of possible boundary conditions, for
each of which at least one of the 2-spinor components must be irregular at
$r=0$.

Mathematically, this corresponds to the fact that the Hamiltonians
$H_n^{\uparrow\downarrow}$ derived from~(\ref{eq:coup}) are essentially
self-adjoint\footnote{An operator $A$ on some domain is {\em essentially
self-adjoint} if its operator closure $\bar{A}$ is self-adjoint i.e. $\bar{A}$
and its adjoint $\bar{A}^*$ have the same domain on which they act in the same
way.} \cite{RSi} on $C^{\infty}$-spinors compactly supported away from $r=0$
in $L^2((0,\infty)^2,rdr)$ in all sectors other than the critical sector where
$H_n^{\uparrow\downarrow}$ have deficiency indices $\langle 1,1\rangle$
\cite{dSG,FK}. This entails that there is a 1-parameter family of self-adjoint
extensions labelled by elements of $U(1)$, each of which corresponds to a
different choice of boundary condition in the critical sector. The
self-adjointness of the Hamiltonian is necessary to ensure a unitary time
evolution, so it is only by choosing a particular self-adjoint extension that
we can specify a well-defined global dynamics for the system (see also
\cite{KF,FK}).

Following the procedure for computing amplification factors introduced in
\cite{Anne} (see equation~(\ref{eq:amplif})), we now impose scattering boundary
conditions on the wavefunction in each sector and then compare the magnitude of
the wavefunction at $r=a$ against that of the free wavefunction, which is of
order 1. The scattering boundary conditions are derived in Appendix A and give
the normalised spinor as:
\begin{eqnarray}
\psi_n & = &
\left(\begin{array}{c}
\left[1-(-i)^{2|n+\Phi|}\tan\theta_n\right]^{-1}
   (-i)^{|n+\Phi|}J_{|n+\Phi|}(kr) \\
\left[1-(-i)^{2|n+\Phi+1|}\tan\varphi_n\right]^{-1}
   \Lambda (-i)^{|n+\Phi+1|}J_{|n+\Phi+1|}(kr)
     \end{array}\right) + \nonumber \\
 & & \left(\begin{array}{c}
\left[1-(-i)^{-2|n+\Phi|}\cot\theta_n\right]^{-1}
   (-i)^{-|n+\Phi|}J_{-|n+\Phi|}(kr)  \\
\left[1-(-i)^{-2|n+\Phi+1|}\cot\varphi_n\right]^{-1}
   \Lambda (-i)^{-|n+\Phi+1|}J_{-|n+\Phi+1|}(kr)
      \end{array} \right)                             \label{eq:spnr}
\end{eqnarray}
where $\Lambda=-k/(\omega+m)$ which is of order 1 at the energies of
interest (where $k\sim\tilde{m}$), for either choice $m=\pm\tilde{m}$.

\section{Scattering Length Formalism}

  In order to determine the wavefunction amplification, it now suffices to
specify $\theta_n$ and $\varphi_n$. This is accomplished by performing the
analogue of the low energy expansion in potential scattering theory. If we
denote the logarithmic derivative $F_{n,k}^{\prime}/F_{n,k}|_{r=a}$ by $D_n$,
we may expand $D_n^+$ in powers of $(ka)^2$:
$D_n^+=D_n^{(0)+}+(ka)^2D_n^{(1)+}+O(ka)^4$. Note that $ka$, the product of
fermion momentum and string radius is of order $10^{-15}$. The matching between
internal and external solutions
\begin{equation}
\cot\theta_n=\frac{D_nJ_{-|n+\Phi|}(kr)-J_{-|n+\Phi|}^{\prime}(kr)}
                  {D_nJ_{|n+\Phi|}(kr)-J_{|n+\Phi|}^{\prime}(kr)}
\end{equation}
(where the prime denotes differentiation with respect to $r$) may be expanded
to give
\begin{equation}
\cot\theta_n=\left(\frac{kR_n^+}{2}\right)^{-2|n+\Phi|}
             \frac{\Gamma(1+|n+\Phi|)}{\Gamma(1-|n+\Phi|)}
             \left\{ 1 -\frac{|n+\Phi|\pi}{2}(kr_n^+)^2+ O(k)^4 \right\}
\label{eq:er1}
\end{equation}
where we call $R_n^+$ the scattering length of the effective potential $V_n^+$
given by
\begin{equation}
R_n^+=a\left[ \frac{aD_n^{(0)+}-|n+\Phi|}
                   {aD_n^{(0)+}+|n+\Phi|}\right]^{1/(2|n+\Phi|)}
\label{eq:fit0}
\end{equation}
and $r_n^+$, which generalises the effective range of potential scattering,
is given by
\begin{eqnarray}
(r_n^+)^2 & = & \frac{a^2}{2\pi|n+\Phi|^2}
       \left[ \frac{(a/R_n^+)^{-2|n+\Phi|}}{1-|n+\Phi|} - 2
            + \frac{(a/R_n^+)^{2|n+\Phi|}}{1+|n+\Phi|}  \right.  \nonumber \\
    &   & \left. +2aD_n^{(1)+}\left( \left(\frac{a}{R_n^+}\right)^{|n+\Phi|}
                 -\left(\frac{a}{R_n^+}\right)^{-|n+\Phi|}\right)^2\right].
\label{eq:r1}
\end{eqnarray}
Our parameter $\theta_n$ is related to the phase shift $\delta_n$ relative to
the Aharonov-Bohm scattering (defined by $F_{n,k}\propto \cos\delta_n
J_{|n+\Phi|}(kr)- \sin\delta_n N_{|n+\Phi|}(kr)$) by
\begin{equation}
\cot\delta_n = \frac{\cos (|n+\Phi|\pi) - \cot\theta_n}{\sin(|n+\Phi|\pi)}.
\end{equation}

Note that there is another definition of scattering length (which is the one
used in \cite{KF,FK}): equation~(\ref{eq:uptrue}) may be solved exactly at
$\omega^2=m^2$ for $r>a$ (where $V_n(r)=(n+\Phi)^2/r^2$) and takes the simple
form
\begin{equation}
F_n(r)\propto \left(\frac{r}{R_n^+}\right)^{|n+\Phi|} -
\left(\frac{r}{R_n^+}\right)^{-|n+\Phi|}
\label{eq:soln}
\end{equation}
which we can use to define $R_n^+$. Note that (to ensure reality of $F_n(r)$ up
to a phase) $(R_n^+)^{2|n+\Phi|}$ must be real (although possibly negative) and
so the allowed values of $R_n^+$ lie on a contour in the complex plane. We make
this definition of $R_n^+$ to ensure that our `scattering length' really has
dimensions of length; it will turn out that this is a natural parametrisation
to use. $R_n^+$ may be expressed in terms of $F_n(r)$ by means of a {\it
fitting formula} \cite{FK}
\begin{equation}
R_n^+ = a \left\{ 1 - \left.
\frac{2|n+\Phi|}{a}\frac{r^{|n+\Phi|}F_n}{(r^{|n+\Phi|}F_n)^{\prime}}
\right|_{r=a}\right\}^{1/(2|n+\Phi|)}
\label{eq:fit1}
\end{equation}
Substituting $D_n^{+(0)}=F_n^{\prime}(a)/F_n(a)$ in~(\ref{eq:fit1}) reduces it
to~(\ref{eq:fit0}) and we see that our two definitions of scattering length
agree.

Similar considerations for $\cot\varphi_n$ (when the logarithmic derivative
$D_n^-=G_n^{\prime}/G_n|_{r=a}$ is expanded as
$D_n^-=D_n^{(0)-}+(ka)^2D_n^{(1)-}+O(ka)^4$) give
\begin{equation}
\cot\varphi_n=\left(\frac{kR_n^-}{2}\right)^{-2|n+\Phi+1|}
             \frac{\Gamma(1+|n+\Phi+1|)}{\Gamma(1-|n+\Phi+1|)}
             \left\{ 1 - \frac{|n+\Phi+1|\pi}{2}(kr_n^-)^2 + O(k)^4 \right\}
\label{eq:er2}
\end{equation}
where $R_n^-$ is determined from the zero energy external solution $G_n(r)$ by
the fitting formula
\begin{equation}
R_n^-  =  a \left\{ 1 - \left.
\frac{2|n+\Phi+1|}{a}\frac{r^{|n+\Phi+1|}G_n}{(r^{|n+\Phi+1|}G_n)^{\prime}}
\right|_{r=a}\right\}^{1/(2|n+\Phi+1|)}   \label{eq:fit2}
\end{equation}
and $r_n^-$ is given by
\begin{eqnarray}
(r_n^-)^2 & = & \frac{a^2}{2\pi|n+\Phi+1|^2}
   \left[ \frac{(a/R_n^-)^{-2|n+\Phi+1|}}{1-|n+\Phi+1|} - 2
         +\frac{(a/R_n^-)^{2|n+\Phi+1|}}{1+|n+\Phi+1|} \right.  \nonumber  \\
    &   & \left. +2aD_n^{(1)-}\left( \left(\frac{a}{R_n^-}\right)^{|n+\Phi+1|}
                 -\left(\frac{a}{R_n^-}\right)^{-|n+\Phi+1|}\right)^2\right].
\label{eq:r2}
\end{eqnarray}
In this case, the zero energy external solution is
\begin{equation}
G_n(r)\propto \left(\frac{r}{R_n^-}\right)^{|n+\Phi+1|} -
\left(\frac{r}{R_n^-}\right)^{-|n+\Phi+1|}
\end{equation}

The remarkable feature of equations~(\ref{eq:uptrue}) and~(\ref{eq:dntrue}) to
which we referred in the introduction is that they may be solved analytically
at zero kinetic energy ($\omega^2=\tilde{m}^2$). This is because they factorise
at zero kinetic energy (see Appendix B) and may be viewed as a consequence of
an abstract supersymmetry possessed by Dirac operators coupled to external
magnetic fields (see \cite{Thall}). In Appendix B, we derive these solutions
and place bounds on their corresponding scattering lengths sector by sector. We
summarise our results in Table 1. In each sector, one of the two scattering
lengths $R_n^{\pm}$ is either zero or infinite and so the corresponding low
energy expansion~((\ref{eq:er1}) or~(\ref{eq:er2})) breaks down. One could
derive the form of the expansion in these special cases; however, the other low
energy expansion remains well-defined and so, by using the formulae connecting
$\theta_n$ and $\varphi_n$, we can always determine both as functions of $k$ at
low energies.

In sectors where the bounds derived in Appendix B allow a range of possible
scattering lengths, the precise details of the flux distribution fix a
particular choice via the fitting formulae. The bounds derived in Appendix B
can be shown to be `best possible' and so, in sectors where our bounds permit,
$R_n^{\pm}$ can be made arbitrarily large for arbitrarily small values of $a$.
This persists in the limit as $a\rightarrow 0$ contradicting the results of
\cite{Hagen}.

Large scattering lengths occur only when the corresponding effective potential
exhibits a potential well, as may be seen by the following argument. If the
effective potential in, say, equation~(\ref{eq:uptrue}) is everywhere
non-negative, then a convexity argument applied to the differential equation
(see \cite{FK}) shows that the zero energy solution $F_n(r)\not\equiv 0$ with
regular boundary conditions at the origin must satisfy
$F_n^{\prime}(a)/F_n(a)\ge 0$. Inserting this in the appropriate fitting
formula~(\ref{eq:fit1}) we find that the corresponding scattering length
satisfies $|R_n^+|<a$. Thus to generate large scattering lengths it is
necessary for the effective potential to exhibit a well. Note that if $\Phi>0$,
it is necessary that the magnetic field $B(r)$ change sign within the core of
the string for $R_n^+$ to be large in the critical sector; for the existence of
a well in $V_n^+$ implies that $\alpha^{\prime}(r)/r=B(r)$ must be negative in
some interval in $(0,a)$. Similarly, for $\Phi<0$, $R_n^-$ is large only if
$B(r)$ changes sign. We have already observed that such configurations are not
necessarily unstable when an effective gauge potential is considered.

Of particular interest in Table 1 are the cases $-[\Phi]\le n\le -1$ for
$\Phi>1$ and $0\le -2-[\Phi]$ for $\Phi<-1$. The infinite scattering lengths in
these sectors are due to the presence of bound states of zero kinetic energy
located at $\omega=-\epsilon(\Phi)m$ (where $\epsilon(x)=\pm 1$ as $x$ is
greater than or less than 0). In accordance with a theorem of Aharonov and
Casher \cite{AC}, there are precisely $[\Phi]$ such states for $\Phi>0$ and
$-[\Phi]-1$ such if $\Phi<0$. Although one of the scattering lengths in the
critical sector $n=-1-[\Phi]$ is infinite, this is not a bound state, as the
wavefunction at zero kinetic energy fails to be square integrable at infinity.
(Recall: $[\Phi]$ is the greatest integer {\it strictly} less than $\Phi$.)

\section{Calculation of Amplification Factors}

  The information in Table 1 and expansions~(\ref{eq:er1}) and~(\ref{eq:er2})
enables us to gain some insight into the form of the wavefunction amplification
at low energies (in comparison with the GUT scale), as we now have some control
over the leading order behaviour of $\cot\theta_n$ and $\cot\varphi_n$. We can
use this to construct simple order of magnitude arguments which are sufficient
to demonstrate the range of possible behaviour. In particular, we will assume
that $aD_n^{(1)\pm}$ have magnitude of order 1 or smaller -- we will not
consider the additional behaviour occurring if $D_n^{(1)\pm}$ are finely tuned
so as to produce cancellations in~(\ref{eq:r1}) and~(\ref{eq:r2}). We also
ignore the effect of higher terms in the low energy expansions~(\ref{eq:er1})
and~(\ref{eq:er2}). These assumptions amount to the approximation
\begin{equation}
(kr_n^+)^2 \approx \left\{
\begin{array}{cl} O\left( (ka)^2(a/R_n^+)^{2|n+\Phi|}\right) & R_n^+\ll a \\
                  O\left( (ka)^2\right) & R_n^+\sim a                     \\
                  O\left( (ka)^2(R_n^+/a)^{2|n+\Phi|}\right) & R_n^+\gg a
\end{array} \right. .                                  \label{eq:er3}
\end{equation}
This allows us to conclude that $\cot\theta$ is well-approximated by the first
term in~(\ref{eq:er1}) provided that
\begin{equation}
a(ka)^{1/|n+\Phi|} \ll |R_n^+| \ll a (ka)^{-1/|n+\Phi|}  \label{eq:range}
\end{equation}
and so $\cot\theta_n =O((kR_n^+)^{-2|n+\Phi|})$. In the language of \cite{KF},
we say that $R_n^+$ is `believable' at scale $k^{-1}$. For very small
scattering lengths $R_n^+\ll a(ka)^{1/|n+\Phi|}$, we have $\cot\theta_n\gg
(ka)^{-2(1+|n+\Phi|)}$ and for very large scattering lengths $R_n^+\gg
a(ka)^{-1/|n+\Phi|}$, we have $\cot\theta\sim (ka)^{2(1-|n+\Phi|)}$. Note in
particular that QCD scale scattering lengths $R_n^+=O(k^{-1})$ are classified
as `very large' unless $|n+\Phi|<1$ when they fall within the
range~(\ref{eq:range}). Similar conclusions may be derived for $\cot\varphi$.

  We treat the case $\Phi>0$ and distinguish two cases: the critical sector
$n=-1-[\Phi]$ and other sectors with $n\not=-1-[\Phi]$. In the critical sector,
we find (inserting the coupling relation $\tan\theta_n=-\cot\varphi_n$
in~(\ref{eq:spnr}))
\begin{eqnarray}
\psi_n & = & \left[1-(-i)^{2(1-\nu)}\tan\theta_n\right]^{-1}
\left(\begin{array}{c} (-i)^{1-\nu}J_{1-\nu}(kr) \\
               \Lambda (-i)^{-\nu}J_{-\nu}(kr)
\end{array}\right)   \nonumber \\
     &   & + \left[1-(-i)^{2(\nu-1)}\cot\theta_n\right]^{-1}
\left(\begin{array}{c} (-i)^{\nu-1}J_{\nu-1}(kr) \\
               \Lambda (-i)^{\nu}J_{\nu}(kr)
\end{array}\right)
\end{eqnarray}
and so we see that the wavefunction is always amplified at $r=a$, regardless of
the value of $\theta_n$. However, the degree to which it is amplified is
determined by $\theta_n$ and can vary between $(ka)^{-\nu}$ and $(ka)^{\nu-1}$,
depending on the details of the internal flux distribution. As $|n+\Phi|<1$ in
this sector, we find that we can approximate the expression~(\ref{eq:er1}) for
$\cot\theta_n$ by its first term over a large range of values for $R_n^+$
including QCD length scales. We find that for `very small' scattering lengths
or values of $R_n^+$ within a few orders of magnitude of the tube radius $a$
(i.e. for GUT scale scattering lengths) the amplification is of order
$(ka)^{-\nu}$ and will occur in the lower component of the spinor. For
scattering lengths on the QCD scale, for which $kR_n^+\sim 1$, the
amplification is given by the larger of $(ka)^{-\nu}$ (lower component) and
$(ka)^{\nu-1}$ (upper component), whilst for $R_n^+$ in excess of the QCD
scale, the amplification is of order $(ka)^{\nu-1}$, occurring in the upper
component. Thus the amplification factor is strongly dependent on the details
of the flux distribution.

  In sectors other than $n=-1-[\Phi]$, we find
\begin{eqnarray}
\psi_n & = & \left[1-(-i)^{2|n+\Phi|}\tan\theta_n\right]^{-1}
\left(\begin{array}{c} (-i)^{|n+\Phi|}J_{|n+\Phi|}(kr) \\
               \Lambda (-i)^{|n+\Phi+1|}J_{|n+\Phi+1|}(kr)
\end{array}\right) \nonumber \\
     &   & + \left[1-(-i)^{2|n+\Phi|}\cot\theta_n\right]^{-1}
\left(\begin{array}{c} (-i)^{-|n+\Phi|}J_{-|n+\Phi|}(kr) \\
               \Lambda (-i)^{-|n+\Phi+1|}J_{-|n+\Phi+1|}(kr)
      \end{array}\right).
\end{eqnarray}

  We treat the case $-[\Phi]\le n\le -1$ first which arises only for $\Phi>1$
and corresponds to the Aharonov-Casher states. Here, we have $0<R_n^+< a$. For
cases in which $R_n^+$ is of the order of $a$, we see that
$\cot\theta_n=O((ka)^{-2|n+\Phi|})$ as $(kr_n^+)^2\ll 1$ and so
\begin{equation}
\psi_n\sim \left(\begin{array}{c} O((ka)^{|n+\Phi|}) \\
                                O((ka)^{2|n+\Phi|-|n+\Phi+1|})
                 \end{array}\right)
= \left(\begin{array}{c} O((ka)^{|n+\Phi|}) \\
                                O((ka)^{n+\Phi-1})
                 \end{array}\right)
\end{equation}
and we therefore find amplification only for $n=-[\Phi]$, where the
amplification factor is $(ka)^{\nu-1}$ in the lower component. However, if
$R_n^+$ is `very small' so that $(kr_n^+)^2$ is no longer negligible,
$\cot\theta_n\gg (ka)^{-2(1+|n+\Phi|)}$ and it is easy to see that there is no
amplification even in the sector $n=-[\Phi]$.

  We now treat the case $n\le -2-[\Phi]$. In these sectors it is easy to see,
using the same methods as above, that `very small' scattering lengths or
scattering lengths within a few orders of magnitude of the core radius $a$ give
no amplification. However, if $R_n^+$ is of the order of the QCD scale or
larger, $\cot\theta_n=O((ka)^{2(1-|n+\Phi|)})$, giving
\begin{equation}
\psi_n\sim \left(\begin{array}{c} O((ka)^{|n+\Phi|-2}) \\
                                O((ka)^{2|n+\Phi|-2-|n+\Phi+1|})
                 \end{array}\right)
\end{equation}
and so there is amplification only in the sector $n=-2-[\Phi]$, with
amplification factor $(ka)^{-\nu}$ in the upper component. The possibility of
amplification from this sector does not appear to have been noted before. To
conclude the analysis for $\Phi>0$, it remains to consider $n\ge 0$. In these
sectors, $R_n^+$ is forced to be zero and we must therefore consider $R_n^-$ to
give $\cot\varphi_n$ and then use $\cot\theta_n=-\cot\varphi_n$. As before, we
find no amplification if $R_n^-$ is `very small' or within a few orders of
magnitude of $a$. In the case where $R_n^-$ is QCD scale or larger, we find
$\cot\varphi_n=O((ka)^{2(1-|n+\Phi+1|)})$ and so
\begin{equation}
\psi_n\sim \left(\begin{array}{c} O((ka)^{2|n+\Phi+1|-2-|n+\Phi|}) \\
                                O((ka)^{|n+\Phi+1|-2})
                 \end{array}\right).
\end{equation}
Thus there is no amplification for $n\ge 0$ unless $[\Phi]=0$ in which case,
there is amplification of order $(ka)^{\nu-1}$ in the lower component in sector
$n=0$ only. Note that in this case, there are no Aharonov-Casher states, and
that $n=0$ is adjacent to the critical sector.

  We can derive the analogous results for $\Phi<0$ by sending $\Phi\rightarrow
-\Phi$, $\nu\rightarrow 1-\nu$, $n\rightarrow -1-n$ and $R^+\rightarrow R^-$.
We find that the critical sector $n=-1-[\Phi]$ always exhibits amplification:
for values of $R_n^-$ within a few orders of magnitude of $a$ or smaller,
amplification occurs in the upper component with factor $(ka)^{\nu-1}$, whilst
for $R_n^-$ in excess of the QCD scale, the lower component is amplified by
factor $(ka)^{-\nu}$. In the Aharonov-Casher states $0\le n\le -2-[\Phi]$, we
find that only $n=-2-[\Phi]$ contributes with amplification $(ka)^{-\nu}$ in
the upper component unless $R_n^-\ll a(ka)^{1/|n+\Phi+1|}$ when there is no
amplification. The sector $n=-[\Phi]$ allows amplification $(ka)^{\nu-1}$ of
the lower component only for large $R_n^-$ and there can be amplification in
$n=-1$ only for $[\Phi]=-1$, when the upper component is amplified by
$(ka)^{-\nu}$ only for large $R_n^+$. No other sectors contribute.

  In conclusion, and subject always to the provisos stated before
equation~(\ref{eq:er3}) there are thus at most three sectors which can
contribute to wavefunction amplification: the critical sector $n=-1-[\Phi]$ (in
which there is always amplification) and the two adjacent sectors. We have also
seen that, of the sectors adjacent to the critical sector, only the
Aharonov-Casher sector (when present) amplifies generically and requires an
anomalously small scattering length to be suppressed, while non-Aharonov-Casher
sectors require anomalously large scattering lengths in order to contribute. We
summarise our results in Figure 1, where we graph $p(\Phi)$, which determines
the overall amplification factor of equation~(\ref{eq:amplif}) as
$A=(ka)^{-p(\Phi)}$. Figure 1(a) shows the maximum (solid line) and minimum
(dotted line) possible amplification for each flux. In the most general case,
when we make no assumptions about the form of the flux distribution, we can say
no more than this without explicitly computing the relevant scattering lengths.

  However, if we know that the flux distribution is single-signed within the
core, then if $\Phi>0$ we have $V_n^+\ge 0$ for all $n$ and so our earlier
arguments show that $0<R_n^+<a$; conversely, if $\Phi<0$, we know that
$0<R_n^-<a$. In either case, (and provided $|\Phi|>1$) the relevant
scattering lengths in the contributing sectors are bounded between $0$ and $a$.
(In the case $|\Phi|<1$ (in which there are no Aharonov-Casher states), there
is a possible contribution in the $n=0$ ($n=-1$) sector for $\Phi>0$ ($\Phi<0$)
which is due to $R_0^-$ ($R_{-1}^+$) and therefore unaffected by these bounds.)
Thus for a sufficiently `nice' subclass of single-signed flux distributions,
the scattering lengths of interest will all be within a few orders of magnitude
of $a$ (i.e. of the GUT scale) and so $p(\Phi)$ follows the graph shown in Fig.
1(b) as only the critical sector and (if $|\Phi|>1$) the adjacent
Aharonov-Casher sector contribute. The solid lines indicate the range of $\Phi$
for which the critical sector provides the dominant contribution to the
amplification, whilst the dotted portions indicate the ranges where the
adjacent Aharonov-Casher state gives the dominant amplification. This situation
holds for many simple models of flux distribution (in particular for those
examined in \cite{Anne}) and, as we have indicated, for all sufficiently `nice'
single-signed flux distributions. The results of \cite{Anne} correspond to (and
agree with) our results in this case and so our discussion has demonstrated the
extent to which those results can be considered generic.

We note that for $|\Phi|>1/2$ Fig. 1(b) follows the maximum amplification plot
and so the effect of abnormally large or small scattering lengths could change
the amplification factor only by suppressing it, as would occur, for example,
if the scattering length in the critical sector was large (QCD scale or
larger), or the scattering length in the adjacent Aharonov-Casher sector was
much smaller than $a$. When $|\Phi|<\frac{1}{2}$, however, it is possible to
increase the amplification factor considerably by tuning the scattering length
in the critical sector to be large (a similar effect occurs in this case if the
scattering length in $n=0$ ($\Phi>0$) or $n=-1$ ($\Phi<0$) is tuned to be
large). The special status of $|\Phi|<\frac{1}{2}$ in Fig. 1(b) is due to the
absence of an Aharonov-Casher state in this case, which provides the dominant
amplification when $|\Phi|>1$ and the fractional part of $|\Phi|$ is less than
$\frac{1}{2}$.

  Finally, Fig. 1(c) shows the results which would be obtained using the
thin-wire approximation, on the assumption that the scattering length in the
critical sector (which is the only free parameter) is of the GUT scale. The
inadequacy of this approximation is seen by its disimilarity to Fig. 1(b) and
the importance of the adjacent Aharonov-Casher state becomes clear. It is
important to note that the thin-wire does not support Aharonov-Casher states as
they fail to be normalisable.

\section{Conclusion}

  We first consider the relation of our current results to our other work
\cite{KF,FK} on the large scale effects of small objects. In \cite{KF} we point
out that in many physical situations, a small object may be replaced by a
point-like or line-like idealisation and that if the dynamics of the idealised
system admits more than one consistent choice of boundary condition (in our
case, the Hamiltonian fails to be essentially self-adjoint on a suitable
domain), this is often a signal that the large scale behaviour may be sensitive
to the details of the internal structure of our original small object.
Furthermore, in such cases, the large scale dynamics of the true system is
well-approximated by the idealisation with an appropriate choice of boundary
condition (here a self-adjoint extension) and that therefore the parameter(s)
labelling the choice of boundary condition (in our case, the scattering
lengths)
parametrise the possible large scale behaviour. This is the content of the
``principle of sensitivity'' enunciated in \cite{KF}.

  In the case at hand {\em for the specific purpose of computing wavefunction
amplification factors} the thin wire approximation fails to be a good
idealisation of the true system because there are contributions from
non-critical sectors, generically from the adjacent Aharonov-Casher sector. We
note however, that the Aharonov-Casher states appear for a quite special and
deep reason: an index theorem related to the abstract supersymmetry of the
Dirac operator, and so confirms our general philosophy in \cite{KF} that when
the principle of sensitivity fails to apply, it fails for
`interesting reasons'.

  However, we see that the scattering length formalism developed in
\cite{KF,FK} is still applicable and that the amplification factor in the
critical sector is strongly sensitive to the details of the internal flux
distribution. Moreover, if one considers the elastic scattering cross section
rather than wavefunction amplification factors, it is found that the main
deviation from the pure Aharonov-Bohm cross section at low energies (large
scales) occurs in the critical sector and is parametrised by the scattering
length there. Also, one can show \cite{FK} that if a sequence of Dirac
Hamiltonians describing flux tubes of steadily decreasing radius has a limit
(in a suitable sense of convergent dynamics, technically strong resolvent
convergence) which is self-adjoint (i.e. a well-defined limiting dynamics) then
the limit must be a self-adjoint extension of the idealised thin-wire
approximation. Thus the principle of sensitivity seems to apply as far as
scattering cross sections are concerned.

  We also note that if one modifies the Hilbert space or the domain on which
the Hamiltonian is defined in an appropriate way, it is possible to arrange
that the sectors $n=-2-[\Phi]$, $n=-1-[\Phi]$ and $n=-[\Phi]$ are precisely the
sectors in which the Hamiltonian fails to be essentially self-adjoint and that
therefore it might be that our results can be reconciled with a discussion of
self-adjoint extensions after all. This may be done in a variety of ways; for
example by taking the Hilbert space to be the Sobolev space given by the
completion of the space of smooth spinors compactly supported away from the
flux line in the norm defined by $\langle \phi\mid\psi\rangle=\langle \phi\mid
H^2\psi\rangle_{L^2}$, where $H$ is the thin wire Hamiltonian. Alternatively,
one can keep the original Hilbert space, whilst restricting the domain of the
Hamiltonian to be the range of the massless thin wire Hamiltonian acting on
smooth spinors compactly supported away from the flux line. We hope to return
to the significance of these modified versions of the thin wire approximation
elsewhere.

  We conclude with various remarks. Firstly, the above arguments have
established that the amplification factors for baryon decay enhancement
calculations can depend substantially on the internal distribution of the
magnetic flux. In particular, we note that the case $|\Phi|<\frac{1}{2}$ in
which we have seen that amplification can be increased includes two of the most
physically interesting cases in the GUT model of \cite{AW}, where scattering of
the $d$ quark is modelled by $\Phi=-\frac{1}{4}$, and the electron as
$\Phi=\frac{1}{4}$. For these values of $\Phi$, the difference between
$(ka)^{-\nu}$ and $(ka)^{\nu-1}$ amounts to 8 orders of magnitude. It is
therefore of some importance that the amplification factor be correctly
computed, taking into account the details of the model. The scattering length
formalism presented here provides a convenient calculational scheme.

  Secondly, we turn to the physical interpretation of the process of
wavefunction amplification. From above, it is clear that for a particular
component to be amplified it is necessary that its corresponding (zero energy)
scattering length be large (QCD scale or above). (That it is not sufficient may
be seen by considering the Aharonov-Casher states). We also saw above that
large scattering lengths occur only when the effective potential exhibits a
well of negative potential. This makes it reasonable to suggest that the
physical cause of wavefunction amplification (and therefore of baryon decay
enhancement) is a resonance phenomenon caused by the spin-flux interaction:
incoming quarks may tunnel into the well and be delayed, perhaps being
reflected by the walls of the well before tunnelling out. Quarks are therefore
present in the core of the string much longer than would naively be expected
and therefore decay processes occur with increased probability.

  This interpretation of the enhancement process as a resonance phenomenon
depends on an examination of the details of the effective
potentials~(\ref{eq:Vplu}) and~(\ref{eq:Vmin}) -- in particular the presence or
absence of wells. Thus our interpretation did not emerge clearly from previous
work on this subject, where the restricted range of particular models treated
did not display all of the possible qualitative features discussed above. We
have seen in particular that the thin wire approximation (and therefore an
approach based solely on self-adjoint extensions on the usual domain) is
inadequate for this problem, as we have found possible contributions to
enhancement not only from the critical sector $n=-1-[\Phi]$ (which provides the
only contribution in the thin-wire approximation) but from the two sectors
adjacent to this sector. In particular the adjacent Aharonov-Casher sector
provides the dominant enhancement for certain ranges of $\Phi$ (provided the
associated scattering length is of the order of $a$, which is the case e.g. for
the simple flux distributions models of \cite{Anne}). This relation with the
Aharonov-Casher state has not been noted before.

Finally, we note that although we find one more contributing sector than
\cite{Anne}, the amplification arising from the new sector is at most of the
order of that from the other two sectors. This is in accord with the unitarity
bounds established in \cite{Anne}.

{\it Acknowledgments:} We thank Gary Gibbons, Mark Hindmarsh and Lloyd Alty for
useful conversations and B. Thaller for making a copy of reference \cite{Thall}
available to us in advance of publication. C.J.F. thanks Churchill College,
Cambridge for the award of a Gateway Studentship. B.S.K. thanks SERC for the
award of an Advanced Fellowship and the Schweizerischer Nationalfonds for
partial support. We both thank the Institute for Theoretical Physics at the
University of Berne, Switzerland for hospitality as this work was completed.

\appendix
\section{Scattering Normalisation}

In this Appendix, we derive the scattering boundary conditions required above.
The scattering theory is determined by a relation of form
\begin{equation}
\alpha_n (-i)^{|n+\Phi|}J_{|n+\Phi|}(kr)
+\beta_n (-i)^{-|n+\Phi|}J_{-|n+\Phi|}(kr)
\stackrel{r\rightarrow\infty}{\longrightarrow} (-i)^nJ_n(kr) +
\frac{f_n e^{ikr}}{\sqrt{r}}
\end{equation}
where the $f_n$ are the scattering amplitudes and the integer order Bessel
functions arise from the expansion of the incoming plane wave. All Bessel
functions may be replaced by their asymptotic forms $J_{\mu}(x)\sim
\sqrt{2/(\pi x)}\cos(x-(\mu +\frac{1}{2})\pi/2)$. The scattering normalisation
is then determined by requiring the coefficients of $e^{-ikr}$ to match and
gives $\alpha_n+\beta_n=1$. This leads to the normalised spinor~(\ref{eq:spnr})
given in the text.

\section{Bounds on Scattering Lengths}

  We derive here the range of allowed $R_n^{+}$ sector by sector.
The scattering lengths $R_n^{-}$ may be derived from these by
$R_n^{-}(\Phi)=R_{-1-n}^{+}(-\Phi)$.
Equation~(\ref{eq:uptrue}) at zero kinetic energy ($\omega^2=\tilde{m}^2$) is
\begin{equation}
\left\{\frac{1}{r}\frac{d}{dr}r\frac{d}{dr} -
\left(\frac{n+\alpha(r)}{r}\right)^2+\frac{\alpha^{\prime}(r)}{r}\right\}\psi=0
{}.
\end{equation}
This factorises as
\begin{equation}
\left( \frac{d}{dr}+\frac{n+1+\alpha(r)}{r}\right)
\left( \frac{d}{dr}-\frac{n+\alpha(r)}{r}\right) \psi =0
\end{equation}
and so may be solved exactly to give two independent solutions
\begin{eqnarray}
\psi^{(1)}(r)
& = & r^n\exp\left\{\int_0^r\alpha(r^{\prime})/r^{\prime} dr^{\prime}\right\}
\\
\psi^{(2)}(r)&=& r^n\exp \left\{ \int_0^r
\frac{\alpha(r^{\prime})}{r^{\prime}} dr^{\prime} \right\}
\int_0^r {r^{\prime}}^{-1-2n}\exp \left\{ -2\int_0^{r^{\prime}}
\frac{\alpha(r^{\prime\prime})}{r^{\prime\prime}} dr^{\prime\prime} \right\}
dr^{\prime}.
\end{eqnarray}
The solution $F_n(r)$ to~(\ref{eq:uptrue}) at zero kinetic energy is the
solution with regular boundary conditions at $r=0$. For $n\ge 0$ this is
clearly $\psi^{(1)}(r)$ whilst for $n<0$, $\psi^{(2)}$ is the appropriate
solution.

{\it Case (i): $n\ge 0$} For $r>a$, $\psi^{(1)}(r)\propto r^{n+\Phi}$.
Comparing with~(\ref{eq:soln}) or using the fitting formula~(\ref{eq:fit1}), we
see that $R_n^+=0$ if $n\ge -[\Phi]$ (matching to $r^{|n+\Phi|}$) or
$R_n^+=\infty$ otherwise (matching to $r^{-|n+\Phi|}$).

{\it Case (ii): $n<0$} In general, $\psi^{(2)}(r)$ matches to a non-trivial
linear combination of $r^{\pm |n+\Phi|}$ so the situation is more complex.
The logarithmic derivative is given by
\begin{equation}
\left.\frac{F_n^{\prime}}{F_n}\right|_{r=a} = \frac{n+\Phi}{a}
+\frac{f(a)}{a\bar{f}(a)}
\end{equation}
where $f(r)$ is defined by
\begin{equation}
f(r)=r^{-1-2n}\exp \left\{ -2\int_0^r
\frac{\alpha(r^{\prime})}{r^{\prime}} dr^{\prime} \right\}
\label{eq:f}
\end{equation}
and $\bar{f}(r)$ by
\begin{equation}
\bar{f}(r)=\frac{1}{r}\int_0^r f(r^{\prime})dr^{\prime}
\end{equation}
Clearly, $f(r)$ and $\bar{f}(r)$ are positive and non-vanishing except at
$r=0$. We may rewrite the fitting formula~(\ref{eq:fit1}) as
\begin{equation}
(R_n^+/a)^{2|n+\Phi|} = 1-\frac{2|n+\Phi|}{a}
\left[\frac{|n+\Phi|}{a}+
\left.\frac{F_n^{\prime}}{F_n}\right|_{r=a}\right]^{-1}.
\end{equation}
Thus if $n\ge -[\Phi]$, we obtain
\begin{equation}
R_n^+=a\left[ 1-\left(1+\frac{1}{2|n+\Phi|}\frac{f(a)}{\bar{f}(a)}\right)^{-1}
\right]^{1/(2|n+\Phi|)}
\end{equation}
from which we can conclude the bound $0<R_n^+<a$, given our observations about
$f(r)$ and $\bar{f}(r)$. On the other hand, if $n<-[\Phi]$ we find
\begin{equation}
R_n^+=a\left[ 1-2|n+\Phi|\frac{\bar{f}(a)}{f(a)} \right]^{1/(2|n+\Phi|)}
\end{equation}
yielding the bound $-\infty<(R_n^+)^{2|n+\Phi|}<a^{2|n+\Phi|}$. Note that
$R_n^+$ must be finite as a consequence of the non-vanishing of $f(a)$.

Furthermore, one can show that the above bounds are best possible in the sense
that, for given $n$ and any radius $a$, there exist magnetic flux distributions
supported within radius $a$ with any scattering length in the above allowed
ranges. This may be proven by observing that a potential $V(r)$ takes the
form~(\ref{eq:Vplu}) with $\alpha(r)$ smooth and obeying $\alpha(r)=\lambda r^2
+ O(r^4)$ as $r\rightarrow 0$ and $\alpha(r)=\Phi$ for $r\ge a$, if and only if
the equation
\begin{equation}
\left(-\frac{1}{r}\frac{d}{dr}r\frac{d}{dr} +V\right)u=0
\end{equation}
has a smooth solution $u(r)$ which is non-vanishing in $(0,\infty)$ and obeys
$u(r)=r^n(1+\lambda r^2+O(r^4))$ as $r\rightarrow 0$ and $u(r)\propto
r^{n+\Phi}$ for $r\ge a$, whereupon we may identify
\begin{equation}
\alpha(r)=r\frac{u^{\prime}}{u} - n.
\end{equation}
Full details will appear in \cite{FK}.

We remark that in \cite{Hagen} it is mistakenly concluded that in the limit as
$a\rightarrow 0$, the upper spinor component is always either $J_{|n+\Phi|}$ or
$J_{-|n+\Phi|}$ and that (translated into our language) $R_n^+\rightarrow 0$ if
$\Phi>0$ in the critical sector. However, our arguments above and in \cite{FK}
hold for arbitrarily small $a$ and thus the range of scattering lengths allowed
in the limit as $a\rightarrow 0$ is simply the appropriate limit of the range
for finite $a$. Thus in the critical sector for $\Phi>0$, any scattering length
in the range $-\infty<(R_n^+)^{2|n+\Phi|}\le 0$ is allowed as the limit of
scattering lengths of a sequence of flux tubes of decreasing radius. In
\cite{FK} we also prove rigorous statements about the convergence of the
associated sequence of Hamiltonians.

\newpage
\noindent
\begin{tabular}{||l|l|l||} \hline
& & \\
$n\ge 0$ & $R_n^+=0$ & $-\infty<(R_n^-)^{2|n+\Phi+1|}<a^{2|n+\Phi+1|}$ \\
& & \\ \hline
& & \\
$-[\Phi]\le n\le -1$ & $0<R_n^+<a$ & $R_n^-=\infty$ \\
& & \\ \hline
& & \\
$n=-1-[\Phi]$ & $-\infty<(R_n^+)^{2|n+\Phi|}<a^{2|n+\Phi|}$ & $R_n^-=\infty$ \\
& & \\ \hline
& & \\
$n\le -2-[\Phi]$ & $-\infty<(R_n^+)^{2|n+\Phi|}<a^{2|n+\Phi|}$ & $R_n=0$ \\
& & \\ \hline
\end{tabular}
\vskip 20pt
\center{{\bf Table 1(a):} Allowed scattering lengths for $\Phi>0$.}
\vskip 40pt
\begin{tabular}{||l|l|l||} \hline
& & \\
$n\ge-[\Phi]$ & $R_n^+=0$ & $-\infty<(R_n^-)^{2|n+\Phi+1|}<a^{2|n+\Phi+1|}$ \\
& & \\ \hline
& & \\
$n=-1-[\Phi]$ &$R_n^+=\infty$
& $-\infty<(R_n^-)^{2|n+\Phi+1|}<a^{2|n+\Phi+1|}$\\
& & \\ \hline
& & \\
$0\le n\le -2-[\Phi]$ & $R_n^+=\infty$ & $0<R_n^-<a$ \\
& & \\ \hline
& & \\
$n\le -1$ & $-\infty<(R_n^+)^{2|n+\Phi|}<a^{2|n+\Phi|}$ & $R_n^-=0$ \\
& & \\ \hline
\end{tabular}
\vskip 20pt
\center{{\bf Table 1(b):} Allowed scattering lengths for $\Phi<0$.}
\newpage
\begin{picture}(405,180)(-20,-20)
\put(0,0){\vector(1,0){380}}
\put(180,0){\vector(0,1){150}}
\put(150,140){$p(\Phi)$}\put(390,-2){$\Phi$}
\multiput(0,0)(60,0){7}{\line(0,-1){5}}
\put(-5,-20){-3}\put(55,-20){-2}\put(115,-20){-1}\put(178,-20){0}
\put(238,-20){1}\put(298,-20){2}\put(358,-20){3}
\put(180,120){\line(-1,0){5}}\put(163,116){1}
\thicklines
\multiput(30,60)(60,0){6}{\line(-1,2){30}}
\multiput(30,60)(60,0){6}{\line(1,2){30}}
\multiput(0,0)(60,0){6}{\multiput(-1.5,0)(2,4){15}{.}}
\multiput(60,0)(60,0){6}{\multiput(-1.5,0)(-2,4){15}{.}}
\end{picture}
\center{Fig.1(a) Maximum and minimum amplification factors.}
\begin{picture}(405,180)(-20,-20)
\put(0,0){\vector(1,0){380}}
\put(180,0){\vector(0,1){150}}
\put(150,140){$p(\Phi)$}\put(390,-2){$\Phi$}
\multiput(0,0)(60,0){7}{\line(0,-1){5}}
\put(-5,-20){-3}\put(55,-20){-2}\put(115,-20){-1}\put(178,-20){0}
\put(238,-20){1}\put(298,-20){2}\put(358,-20){3}
\put(180,120){\line(-1,0){5}}\put(163,116){1}
\thicklines
\put(180,0){\line(-1,2){60}}
\put(180,0){\line(1,2){60}}
\multiput(270,60)(60,0){2}{\line(1,2){30}}
\multiput(30,60)(60,0){2}{\line(-1,2){30}}
\multiput(30,60)(60,0){2}{\multiput(-1.5,0)(2,4){15}{.}}
\multiput(270,60)(60,0){2}{\multiput(-1.5,0)(-2,4){15}{.}}
\end{picture}
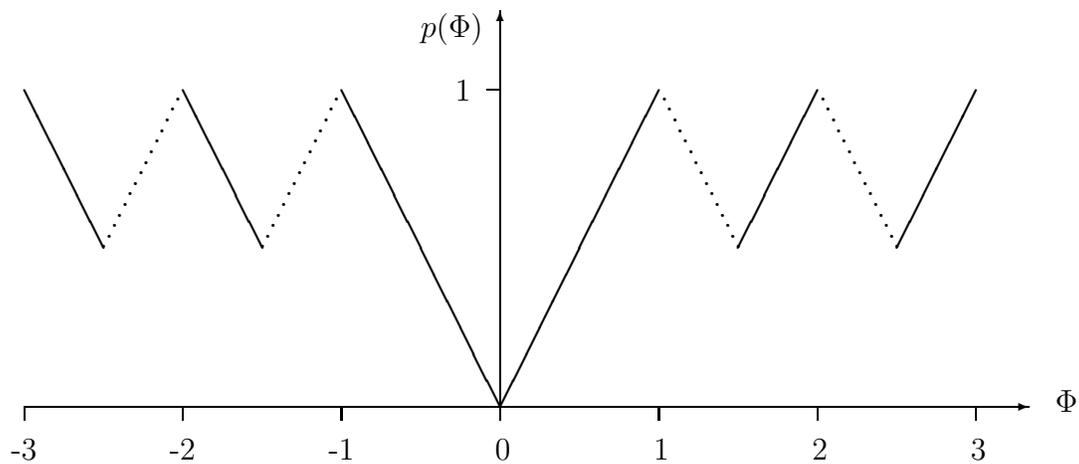
\center{Fig.1(b) Amplification assuming GUT scale scattering lengths.}
\begin{picture}(405,180)(-20,-20)
\put(0,0){\vector(1,0){380}}
\put(180,0){\vector(0,1){150}}
\put(150,140){$p(\Phi)$}\put(390,-2){$\Phi$}
\multiput(0,0)(60,0){7}{\line(0,-1){5}}
\put(-5,-20){-3}\put(55,-20){-2}\put(115,-20){-1}\put(178,-20){0}
\put(238,-20){1}\put(298,-20){2}\put(358,-20){3}
\put(180,120){\line(-1,0){5}}\put(163,116){1}
\thicklines
\multiput(60,0)(60,0){3}{\line(-1,2){60}}
\multiput(180,0)(60,0){3}{\line(1,2){60}}
\end{picture}
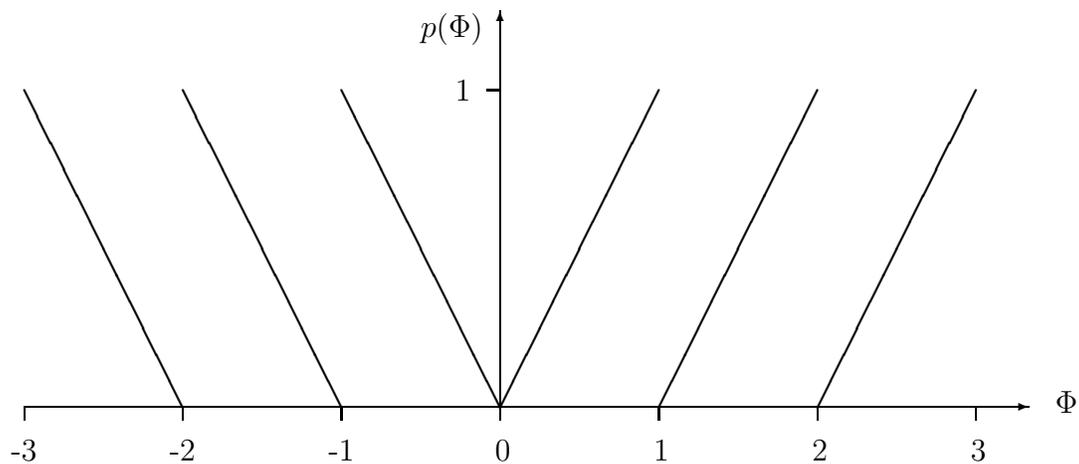
\center{Fig.1(c) Amplification from the critical sector with GUT scale
scattering lengths.}

\end{document}